\documentclass[10pt]{sig-alternate-05-2015}
\usepackage{url}
\usepackage{graphicx}
\usepackage{xspace}
\usepackage{listings}
\usepackage{fancyvrb}
\usepackage[final]{changes}

\newcommand{\yrp}{Yarrp\xspace}
\newcommand{\eg}{e.g.\ }
\newcommand{\ie}{i.e.\ }
\newcommand{\etal}{et al.\ }
\providecommand{\e}[1]{\ensuremath{\times 10^{#1}}}

\begin{document}

\CopyrightYear{2016}
\setcopyright{usgov}
\conferenceinfo{IMC 2016,}{November 14-16, 2016, Santa Monica, CA, USA}
\isbn{978-1-4503-4526-2/16/11}\acmPrice{\$15.00}
\doi{http://dx.doi.org/10.1145/2987443.2987479}

\clubpenalty=10000 
\widowpenalty = 10000

\title{Yarrp'ing the Internet: Randomized High-Speed
Active Topology Discovery}

\numberofauthors{1}
\author{
 \alignauthor Robert Beverly \\
 \affaddr{Naval Postgraduate School}\\
 \email{rbeverly@nps.edu}
}

\maketitle
\thispagestyle{empty}


\begin{abstract}
Obtaining a ``snapshot'' of the Internet topology remains
an elusive task.  Existing active topology discovery techniques
and \deleted{production }systems require significant probing time -- time during
which the underlying network may \deleted{change or }experience
\replaced{transient}{short-lived}
dynamics.
\replaced{This work considers}{In this work, we consider} how active probing can
gather the
Internet topology in \emph{minutes} rather than days.
Conventional approaches to active topology mapping face two
primary speed and scale impediments: i) per-trace state maintenance;
and ii) a low-degree of parallelism.
Based on this observation, we develop
\yrp (Yelling at Random
Routers Progressively), a new traceroute technique designed for high-rate,
Internet-scale probing.  
\yrp is stateless, reconstituting all
necessary information from ICMP replies as they arrive asynchronously.  To avoid 
overloading routers or links with probe traffic, \yrp randomly
permutes an input $IP \times TTL$ space.
We run \yrp at $100Kpps$, a rate at which the paths to all IPv4 /24's
can
be mapped in approximately one hour from a single vantage point.
We compare \yrp\deleted{'s topological recall and discovery rate }against 
existing systems, and present \replaced{examples}{some of the first
results} of
topological dynamics exposed via the high sampling rates
\yrp enables.
\end{abstract}


\section{Introduction}

Network and security researchers rely on topological maps of the
logical Internet to address problems ranging from
critical infrastructure protection to policy.  Production active
measurement systems that continually gather and curate Internet
topology, \eg\cite{iplane06,caida-ark}, are thus important to many longitudinal analyses and to
shedding light on network events of interest. 

Obtaining IP, router, and provider-level network topologies has been a
continual research focus for more than two decades
\cite{huffaker2012topology, claffy09,
spring2002measuring,iplane06}.
While significant progress has been made, topology mapping at Internet
scale remains a challenge.  Both the accuracy of the inferred network
topologies \cite{mathinternet}, and the speed at which they can be
recovered \cite{donnet2005efficient, Beverly:2010:PAI:1879141.1879162}, present
obstacles to current mapping efforts.  
\added{Although systems that
sample paths at a rate inversely proportional to their stability
have shown promise, even state-of-the-art techniques to predict
path changes are relatively inaccurate \cite{cunha2011predicting}.}
In this work, we focus on 
the \emph{speed} and \emph{scale} of Internet-wide active topology
mapping.

Given its scale, experience and popular belief dictates that
obtaining even partial Internet topologies
via active network
probing is a time-intensive process.  For instance, CAIDA's
Archipelago (Ark) \cite{caida-ark} system uses dozens of vantage
points and at least a day to traceroute to a single address in each
routed /24 IPv4 prefix.  
A recent topology cycle gathered by Ark from April, 2016
\cite{caida-topo} sent approximately 11M traceroutes
from 37 monitors over the course of 31 hours in order to discover
$\approx$1M distinct router interfaces, and $\approx$2M
links.

We re-examine \deleted{some of }the assumed fundamental limits of active
topology mapping to consider whether \deleted{such }probing
\replaced{can}{could} be performed
in minutes rather than hours.  Taking inspiration from recent
stateless and randomized high-speed \deleted{network }scanners,
\eg ZMap \cite{durumeric2013zmap} and
masscan \cite{graham14}, we create \yrp (Yelling at Random
Routers Progressively).

To facilitate high-probing rates, \yrp is stateless, reconstituting all
necessary information from replies as they arrive asynchronously.  To avoid 
overloading routers or links, \yrp randomly
permutes its input $IP\times TTL$ space when probing.  \yrp is thus
well-suited for Internet-scale studies.
Our contributions include:
\vspace{-2mm}
\begin{enumerate}
\addtolength{\itemsep}{-0.5\baselineskip}
  \item Development of \yrp, a publicly available tool \cite{yarrp} that  
  permits rapid active network topology discovery. 
  We run \yrp at 
  $100Kpps$ to discover more than 400,000 router interfaces in under 30 minutes.
  \item A comparison of \yrp and CAIDA's existing production topology
   collection platform, showing recall and speed differences.
  \item As an application of rapid topology discovery, we conduct successive topology snapshots 
   separated by a small time delta and characterize the distribution
   and causes of observed path differences.
\end{enumerate}

\section{Background}

Traditional traceroute \cite{traceroute} obtains the sequence of router interface IP
addresses along the forward path to a destination by sending probe
packets with varying time to live (TTL) values and examining the ICMP
responses.  By maintaining the transmission timestamp of each probe, 
traceroute can report the round trip time (RTT)
from the source to each responsive hop.  Modern traceroute
implementations send batches of concurrent probes to lower
tracing time, \eg Linux defaults to 16 simultaneous probes.  In order to match probes
to the ICMP TTL exceeded responses they generate, the probe must include unique
identifiers that are returned as part of the ICMP quotation
\cite{malone07}.  Because the quote is only required to copy the first
28 bytes of the packet that induced the expiry message
\cite{rfc1812},
traceroute typically relies on the first 8 bytes of the
transport-layer header to match responses to probes.  

While various improvements have been proposed and implemented, the
core behavior of traceroute -- and large scale active topology
scanning
-- remains largely unchanged.  To prevent false inferences due to
load-balanced paths, Augustin \etal created Paris traceroute
\cite{augustin2006avoiding}.  To reduce unnecessary probing, Donnet
\etal developed Doubletree \cite{donnet2005efficient}, a modified traceroute that
begins probing from a likely path midpoint outward until it reaches
previously discovered hops.  Similarly, 
\cite{Beverly:2010:PAI:1879141.1879162, baltra14ips}
proposed several
topology primitives empirically shown to reduce the volume of probing
while maintaining or increasing topological discovery.  
\added{DTrack \cite{cunha2011predicting} and Sibyl
\cite{cunha2016sibyl} seek to optimize active probing by making
predictions over historical measurements, yet are still constrained
by traditional traceroute techniques.}

CAIDA's production Ark
infrastructure~\cite{caida-ark} uses
Scamper \cite{luckie10scamper} to perform continual
Internet-wide probing~\cite{caida-topo}.  
Scamper implements both Doubletree and Paris traceroute, has
an open API, can maintain a configurable probing rate, and can be
controlled remotely.  

Traceroute was originally designed as a tool for 
network administrators to diagnose a small number of
paths, not as a means to gather snapshots of the entire Internet
topology \cite{traceroute, claffy09}.
Fundamentally, traceroute and its variants all have 
two \added{related} properties that limit their scalability and speed. They:
\vspace{-1mm}
\begin{itemize}
\addtolength{\itemsep}{-0.5\baselineskip}
  \item Maintain state for each outstanding probe sent,
        including some identifier and origination time.  
  \item Are sequential, probing all hops along the path to a
        destination.  While some tools (\eg scamper) can traceroute
        to multiple
        targets, this parallelism is \replaced{path specific}
        {limited to a finite window of destinations}.
\end{itemize}

In contrast, \yrp is designed to be stateless and random -- probing different
portions of many different paths simultaneously.  This allows \yrp to
send probes at a high per-packet rate, while spreading the load among
many destination networks to avoid concentrating load on particular
paths, links or routers, thereby avoiding anomaly alarms or ICMP
rate limiting.

\section{Yarrp Design}
\label{sec:design}

The high-level idea of \yrp is: 
i) randomization of the probing order of
the domain of network range(s) and TTLs; and
ii) stateless operation, whereby all
necessary state is encoded into the probes such that it can be recovered from
the ICMP replies.  \added{As with ZMap, \yrp uses independent send and
receive threads, where the sender uses raw sockets while the receiver 
thread is implemented using libpcap.}
\yrp is written in C++\added{ (approximately 2,500 SLOC)}, is portable to a variety of UNIX-like
platforms, and is publicly available~\cite{yarrp}.

\subsection{Pseudo-random Probing Order}
\label{sec:design:random}

Existing traceroute techniques probe all hops along a path to a 
destination in sequence.
Instead, we employ a keyed block cipher to provide a bijection over the input
domain of target IPs and TTLs ($D = IPs\times TTLs$).  This
means\added{, for example,}
that \yrp \replaced{may}{will} send a probe to IP address $A$ with $TTL=12$, then
\deleted{toward} $B$ with $TTL=3$, then $C$ with $TTL=20$, and so on until the
entire space of $TTLs$ for each target $IP$ is covered.  \added{To
an outside observer, the probed addresses appear to be random.}

The symmetric RC5 block cipher with a 32-bit block size 
is fast and a natural fit for our application\footnote{Other block ciphers
could be used; the cryptographic strength of the cipher
is not critical to our application.}.  With key $k$, \yrp encrypts the sequence
$i=0,1,\ldots,|D|-1$ where bits of each ciphertext $C_i=RC5_k(i)$
determine the target IP address and TTL to probe.  In this way,
\yrp randomizes the order of probed $\langle target IP, TTL\rangle$.  
\yrp can permute arbitrarily large or small IPv4 address and TTL
domains, or can permute the order of specific targets read from a
file.
Depending on the size of the domain, we switch between
either a prefix-cipher or cycle-walking cipher, as described in
\cite{ciphers}.  

To facilitate comparison with CAIDA's IPv4 topology
dataset~\cite{caida-topo}, \yrp has a mode that
probes a random address in each IPv4 24-bit subnet -- this
mimics the targets selected in a full cycle of CAIDA's probing. 
Here, \yrp encrypts each $i=0,\ldots,2^{24}-1$ with key $k$.
For $C_i = RC5_k(i)$, \yrp probes the IPv4 address  
\replaced{
$C_i[0\mathbin{:}23]*2^8 + (C_i[0:7]+C_i[8:15]+C_i[16:23]) \% 256$ 
}{
$C_i[0\mathbin{:}23]*2^8 + (C_i[0]+C_i[1]+C_i[2]) \% 256$ 
}
with TTL $C_i[24\mathbin{:}31]$.  In this fashion,
we permute through the space of $2^{24}$ possible /24s, and 
construct the least-significant octet as a function of the subnet
such that
the same random address in each /24 is used as the
destination for each TTL.

An advantage of \yrp's randomization method is that the
probing work can easily be distributed among multiple vantage points
with negligible coordination or communication overhead.
We discuss distributed \yrp as
a future enhancement in~\S\ref{sec:conclusions}.

\subsection{Stateless Operation}
\label{sec:design:stateless}

Existing traceroute techniques require state to match ICMP replies
to probes.  In contrast, \yrp does not require state.
We overload various fields in
the probe packets with specific values such that we can reconstruct the corresponding
probe's destination, transmission time, and originating TTL from
within the
quote of the ICMP TTL exceeded messages.  

\begin{figure}[!t]
  \centering
  \resizebox{1.0\columnwidth}{!}{\includegraphics{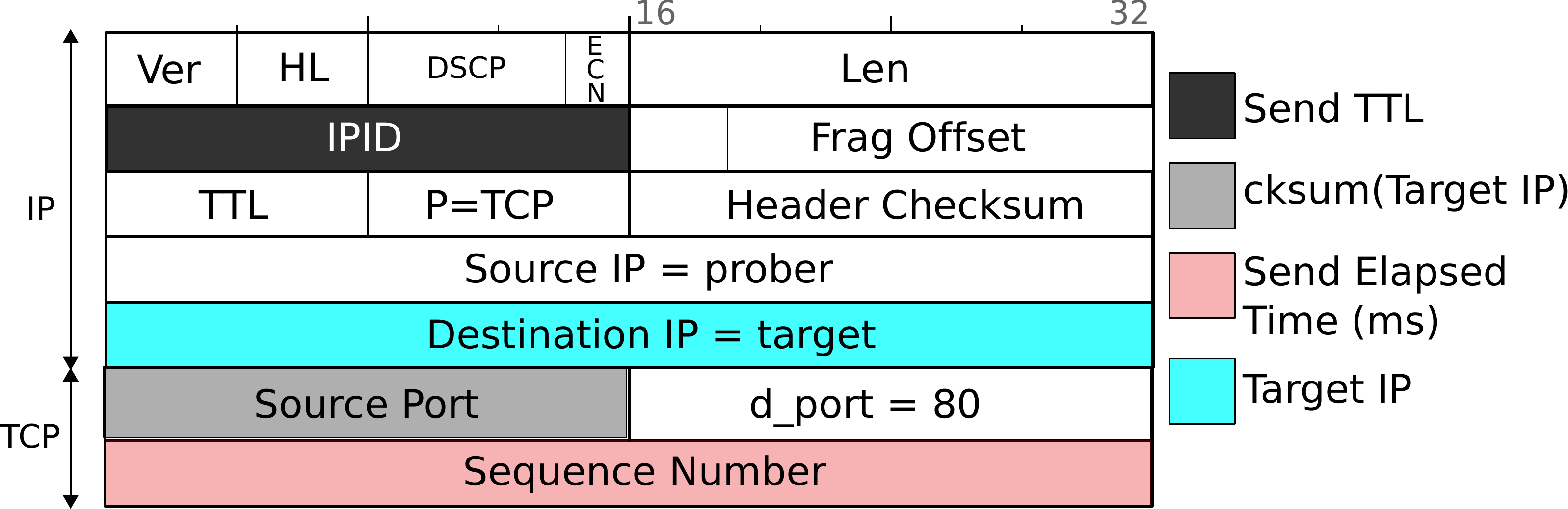}}
  \vspace{-7mm}
  \caption{\yrp encodes information in the IP and TCP fields of
  outgoing probe packets in order to permit stateless operation.}
  \label{fig:fields}
  \vspace{-4mm}
\end{figure}

Figure~\ref{fig:fields} depicts the TCP/IP header fields we utilize.
We encode the TTL with which the packet was sent in the IPID
and the elapsed time in the TCP sequence
number\added{\footnote{\cite{rfc6864} proposes that IPID only
be used for fragmentation.  Should this
become standardized, \yrp can utilize other fields, e.g.,
encoding TTL via packet lengths.}}.  We use elapsed time
rather than \eg Unix time in order to maintain millisecond resolution
with only a 32-bit field.  \yrp can also encode microsecond
resolution, so long as the expected duration of a probing run is less
than $2^{32}/1\e{6} \approx$4300 seconds.  The destination TCP port is
fixed to port 80 to facilitate firewall traversal, while we populate
the source TCP port with the checksum of the target IP address.  In
this fashion, we can detect instances where the destination IP address
is modified enroute, a phenomenon Malone and Luckie observe in 2\%
of their results~\cite{malone07}. 

In order to properly accommodate load-balanced paths,
which are common in the Internet, we ensure that, for
a given destination, certain fields remain fixed for
all TTL probe values.  For instance, although the 
TCP source port changes, it is a function of the 
destination IP address and therefore will contain
the same value for all probes sent toward the destination.
This design allows us to maintain
the benefits of Paris traceroute~\cite{augustin2006avoiding}.  

When ICMP TTL exceeded messages arrive, we examine the included
quotation to recover the destination probed, the originating TTL
(hop), responding interface at that hop, and compute the RTT
by taking the difference of the packet arrival time and the
probe origination time as encoded in the quoted TCP sequence
number.  These values can be computed from the minimum 28 bytes
of required quotation \cite{rfc1812}.

\yrp can source either TCP SYN or ACK probes.  While
SYN probes can permit middlebox traversal, we use the ACK-only
mode to avoid alarms triggered by large volumes of SYN traffic.
\added{We discuss our use of high-rate TCP ACK probing
in~\S\ref{sec:design:ethics}, and outline a means to use
ICMP and UDP probes in future work in~\S\ref{sec:conclusions}.}

\subsection{Challenges}

The benefits of \yrp's design come with several concomitant
challenges, namely: i) reconstructing the unordered responses into 
paths, ii) knowing when to stop probing, and iii) avoiding unnecessary 
probing.  

In following with \yrp's stateless nature, ICMP responses are decoded
as they arrive and written sequentially to a structured output file.
Each entry in the output file corresponds to an ICMP response.  An
entry includes the target IP address, originating TTL, responding
router interface IP address, RTT, and meta-data such as timestamps, IPID,
response TTL, packet sizes, and DSCP markings.  Because of the
inherently random probing, the entries for each hop along a path to a
given destination will be unordered and intermixed with other
responses in the \yrp output file.  We must therefore reconstruct
complete paths by parsing the entire output file and maintaining state
for each destination.  While this is a memory and time-intensive task,
the key point is that it can be performed \emph{off-line}.  In this
fashion, we decouple probing from path reconstruction to permit the
probing to be as fast as possible.  Included in the \yrp distribution
\cite{yarrp}
is a \texttt{yrp2warts} Python script that performs this off-line
conversion into the standard warts \cite{luckie10scamper} binary trace format.

A practical consequence of \yrp's randomization and lack of state is
that its probing behavior does not depend on the received responses.
Thus, \yrp cannot stop probing once it reaches the destination or
when the path contains a sequence of unresponsive hops (the so-called
``gap limit'').  To better understand the optimal range of TTLs to
probe (from the possible space 1-255), we examine the results from a
complete cycle of Ark probing from January, 2016~\cite{caida-topo}.  We seek to
determine, across each of the 
Ark vantage points, the number of unique
router interfaces discovered at each TTL.  

\begin{figure}[!t]
  \centering
  \resizebox{0.92\columnwidth}{!}{\includegraphics{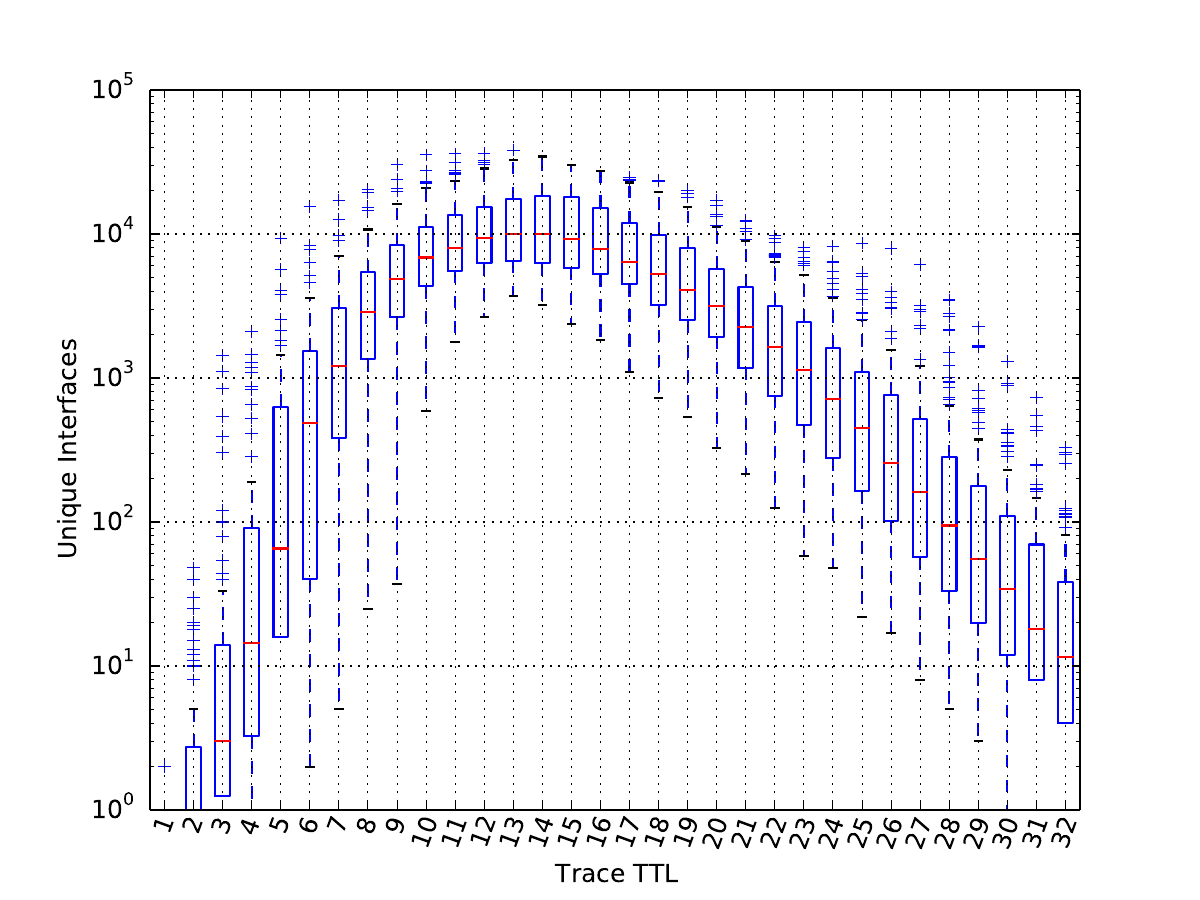}}
  \vspace{-3mm}
  \caption{Distribution of interfaces discovered across Ark vantage points
  in a single cycle as a function of TTL (path depth).  The red line
  indicates the median value among vantage points.}
  \label{fig:ttlhisto}
  \vspace{-3mm}
\end{figure}

Figure~\ref{fig:ttlhisto}
shows the inter-quartile range of the number of distinct interfaces
found as a function of TTL for each of the vantage points; the red
line in the boxplot displays the median number of interfaces per TTL
among each vantage point.  Because of the Internet's tree-like
structure,
the first few hops reveal only a small number of interfaces
regardless of the destination probed.  The bulk of the interfaces are
found between TTLs 10 to 16, with an inflection point around a TTL of
14.  The amount of discoverable topology beyond a TTL of 32 is
negligible (note the log scale y-axis).  As a result, \yrp defaults to
probing TTLs 1 to 32 to minimize unnecessary probing while exploring
the majority of the space.  

For many destinations \yrp will perform more probing
than traditional traceroute methods.  This is both an advantage
and a disadvantage: we show in~\S\ref{sec:results:gaplimit} that
discoverable topology exists beyond multiple unresponsive
hops (where existing methods terminate early).  

\subsection{Optimizations}
\label{sec:design:optimize}

In environments sensitive to probing volume, several
optimizations can substantially decrease unnecessary probing
at the expense of maintaining some state.  This subsection
discusses optimizations to the base \yrp design to enable
different tradeoffs.

First, \yrp can read a BGP routing table of network prefixes 
and build a longest-match Patricia trie~\cite{sklower1991tree}.  
When iterating through the
entire permuted IPv4 space, \yrp can skip destinations that are not
routed.  Based on current global BGP routing tables~\cite{routeviews}, this optimization
avoids probing approximately 1.5B IP addresses (35\% of the 32-bit
space) that are unlikely to return useful results.  Note that the
memory required to maintain the BGP table is constant during a 
probing run (amounting to approximately 300MB during runtime).  In our
experiments, these lookups in the Patricia trie did not prevent
\yrp from running at over $100kpps$.  

Second, the tree-like structure of the network implies that the set of
interfaces near to the vantage point is small relative to the universe
of router interfaces~\cite{donnet2005efficient}.  In
Figure~\ref{fig:ttlhisto} for instance, all of the traces have a single first hop in
common and orders of magnitude fewer interfaces at hops 1-3 as
compared to hops 13-15.  To avoid rediscovering the same nearby router
interfaces repeatedly, \yrp can
maintain state
over the set of responding local ``neighborhood'' interface IP addresses at
hops 1 through a run-time configurable $ttl_{nbrhd}$.  
For each TTL in the neighborhood,
\yrp maintains two timestamps: the last time a probe was sent with
that TTL, and the last time a new interface at that depth replied.  If
no new interfaces have been discovered within the past $\eta=30$
seconds of probing, \yrp skips future probes at that
TTL\added{\footnote{30s ensures $\ge$10 probes
of TTLs 1-8, assuming a balanced binary tree network and 
100Kpps probing rate.  A threshold using the number of
unique interfaces versus probes at a given TTL
may better facilitate adapting to different environments 
without parameterization.}}.  The
\texttt{yrp2warts} script can then stitch together these missing hops.
While the amount of state in neighborhood mode can grown
unbounded, in practice it is small for small $ttl_{nbrhd}$, while
avoiding substantial over-probing.


\subsection{Ethical Concerns}
\label{sec:design:ethics}

\added{
High-speed probing invariably raises ethical concerns, as it increases
the chance that traffic may be perceived as abusive.  We follow the
recommended guidelines for good Internet citizenship provided
in~\cite{durumeric2013zmap} to mitigate the potential impact of our
probing.  
}

\added{
First, as described in \S\ref{sec:design:random}, \yrp's pseudo-random
probing order is designed to avoid overloading the
networks it seeks to characterize.  Second, \yrp sends TCP ACK probes,
which have been used in prior topology studies \cite{keys13},
and prevent end systems from attempting to negotiate a TCP
connection (in contrast, the ZMap scanner sends TCP SYN packets).
Unfortunately, \yrp's stateless nature implies that multiple probes,
with different TTLs, may reach a single destination, an effect we
analyze in \S\ref{sec:results:discovery}.
}

\added{
We therefore make an informative web page accessible via the
IP address of our prober, along with instructions for opting-out.
Additionally, the reverse DNS record name indicates the
research nature of the host.  In this initial work, with \yrp
runs of 30 minutes or less, we did not receive any abuse reports or opt-out
requests.
}

\section{Results}
\label{sec:results}

This section examines results from running \yrp on the Internet.
We compare the topological recall against an existing production
system and then analyze the discovery yield (\ie the amount of new
topology discovered over time).  Finally, as an application of \yrp's
probing speed, we gather \replaced{three}{two} successive
\replaced{topology snapshots}{snapshots of Internet topology
separated in time by a small delta }to reveal instances of short-lived
network dynamics.

%
\subsection{Topological Recall}

We empirically verify \yrp's \replaced{topological recall}{correctness} by evaluating it against
\added{
scamper~\cite{luckie10scamper}.  From a single vantage point, we probe
67,045 destinations using \yrp and scamper.  We run scamper in Paris
TCP ACK mode using port 80 in order to mimic \yrp's behavior and
facilitate an unbiased comparison of topological recall between the
two probing methods.  From the \yrp and scamper probing, construct
graphs of interface
nodes connected by edges when the interfaces appear in consecutive
hops
of a path. We ignore anonymous interfaces
\cite{gunes2008resolving} such that the graph may be disconnected.
\yrp discovers 57,128 interfaces, 1.3\% fewer than scamper's 57,866
unique interfaces, and 67,563 edges (0.8\% more than scamper).  Manual
investigation of the topologies reveals differences mainly
attributable to load-balancing (because scamper and \yrp use different
headers, it is not possible to ensure that they traverse the same
paths to destinations) and path changes.  Empirically, our comparison
demonstrates \yrp's ability to discover
the responsive topology.
}

\deleted{
CAIDA's use of scamper in the Ark infrastructure~\cite{caida-ark, luckie10scamper}.  CAIDA makes
their topology traces publicly available~\cite{caida-topo}.  We
find all 67,045 destinations probed from the San Diego vantage point
on May 1, 2016.  From this same network attachment point, we
instruct
\yrp to probe the same destinations.  In this
fashion, the vantage
point networks and set of destinations are the same,
allowing an unbiased comparison of topological recall between the two
probing methods.

From the \yrp and Ark probing, we construct graphs of interface
nodes connected by edges when the interfaces appear in consecutive hops
of a path. We ignore anonymous interfaces
\cite{gunes2008resolving} such that the graph may be disconnected.
Figure~\ref{fig:wartscmp} displays the resulting graph degree
distributions on a log-log scale.  We observe that the distributions
match closely, empirically supporting \yrp's ability to discover
the responsive topology.

\begin{figure}[!t]
  \centering
  \resizebox{0.9\columnwidth}{!}{\includegraphics{c004710degree.pdf}}
  \vspace{-4mm}
  \caption{Degree distribution of topology gathered by \yrp versus Ark
  when probing the same targets from the same source network.}
  \label{fig:wartscmp}
  \vspace{-3mm}
\end{figure}

However, \yrp discovers 16\% fewer interfaces (80,134 versus
66,939) and 13\% fewer edges (96,763 versus 84,113) than Ark 
in this experiment.
We attribute much of this difference to \yrp's use of 
TCP ACK probes as compared to CAIDA's use of ICMP probes.  In
a survey by 
Key et al., ICMP probes elicit almost 18\% more responses 
as compared to TCP probes~\cite{keys13}.  ICMP and
UDP-based probing are future \yrp enhancements 
(see~\S\ref{sec:conclusions}).
}

\begin{figure}[!t]
  \centering
  \resizebox{0.9\columnwidth}{!}{\includegraphics{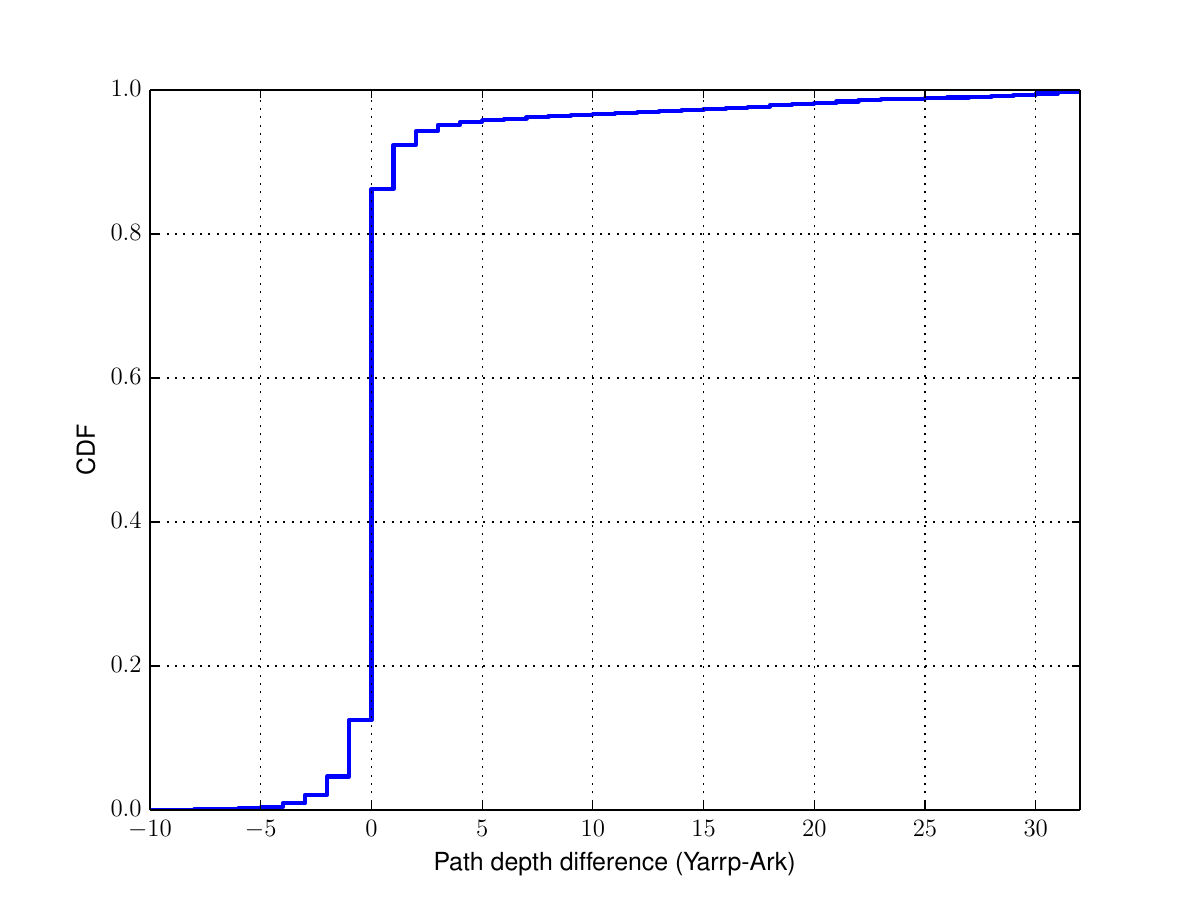}}
  \vspace{-4mm}
  \caption{CDF of highest responding TTL difference
  (max(Yarrp)-max(Ark)) for Ark gap-limited traces (same
  dests. and vantage point).}
  \label{fig:gapdiff}
  \vspace{-3mm}
\end{figure}

\label{sec:results:gaplimit}
\yrp's stateless nature implies that it probes all TTLs
from 1 to 32, whereas Ark's use of scamper ceases probing
after encountering five unresponsive hops in a row.  
In the same CAIDA
San Diego
probing run, 39,613 traces stopped due to this
gap-limit.  For each of these gap-limit traces, we compute
the difference of the highest responding TTL hop from \yrp probing 
and the highest responding TTL from the Ark probing.
Figure~\ref{fig:gapdiff} shows the cumulative distribution of this
difference among the gapped traces; a positive difference means that
\yrp discovered topology beyond the point where Ark stopped probing.
For $\approx$88\% of the
traces, there is no difference.  In 8\% of the traces, Ark
discovers one more hop than \yrp.  However, \yrp discovers one 
additional hop in $\approx6$\% of the targets, and more than 5
additional hops
in 4\% of the cases.   


\subsection{Discovery Rate}
\label{sec:results:discovery}
%

A goal of \yrp is rapid topological discovery.  In this subsection, we
look specifically at the ability to discover unique router interface
addresses rapidly.

On May 10, 2016, we run \yrp from a Northeast United States university vantage point at
$\approx 100kpps$ and instruct it to perform the Ark-mode randomized
probing of the globally routed IPv4 /24 prefixes.  We
limit \yrp to this rate, and limit the duration of our experiment, per
prior agreement with the local network administrator.  The physical
machine is a multi-core Intel L5640 processor running at 2.27GHz, with
\yrp running on an Ubuntu virtual machine allocated a single core.  At
this rate, the CPU utilization is $\approx$52\%.

\begin{figure}[!t]
  \centering
  \resizebox{0.9\columnwidth}{!}{\includegraphics{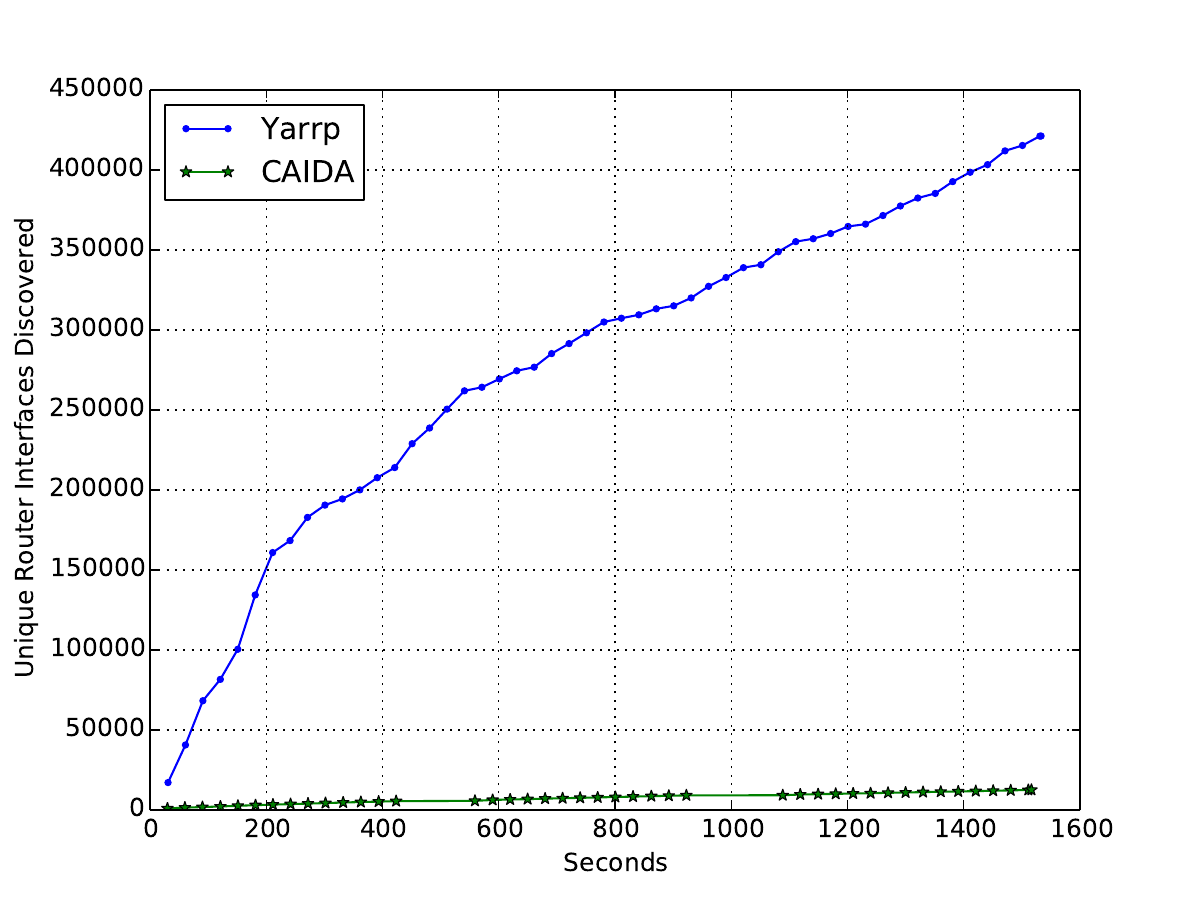}}
  \vspace{-5mm}
  \caption{Unique router interface discovery rate: comparing \yrp
   against CAIDA's routed /24 probing from a single vantage point.}
  \label{fig:discovery}
  \vspace{-3mm}
\end{figure}

We enable the ``neighborhood'' optimization, as described
in~\S\ref{sec:design:optimize}, as we are interested in finding as many
distinct router interfaces as possible given the probing rate.  
Figure~\ref{fig:discovery}
displays the cumulative number of distinct router interfaces
discovered as a function of time.  As a basis of comparison, we also
plot the number of unique interfaces found over time for a single
vantage point (again, using data from the San Diego node
of CAIDA's continual /24 probing~\cite{caida-topo} on May 1, 2016).

CAIDA's San Diego monitor discovers  12,568 unique interfaces in
1,500 seconds ($\approx 8$
per second).  By contrast, \yrp discovers 421,162 unique IPv4
interfaces in the same period, or
approximately 280 distinct router interfaces a second.  The
\added{number of} interfaces found by \yrp in less than 30 minutes equates
to 42\% of \emph{all} unique interfaces discovered from all Ark
monitors over the course of probing for more than a day.

\added{Recall that \yrp decouples probing from topology
reconstruction.  Using a commodity 3.1GHz Intel Xeon processor, our
unoptimized, single-threaded Python program (\texttt{yrp2warts.py})
converts our unordered high-speed \yrp output trace of 6.8M
destinations into an ordered warts-format file in 668 seconds.  This
empirical observation serves to provide an estimate of the wall-clock
upper-bound required to obtain output identical to existing systems.
We leave optimizing \yrp topology reconstruction as future work.}

\begin{figure}[!t]
  \centering
  \resizebox{0.9\columnwidth}{!}{\includegraphics{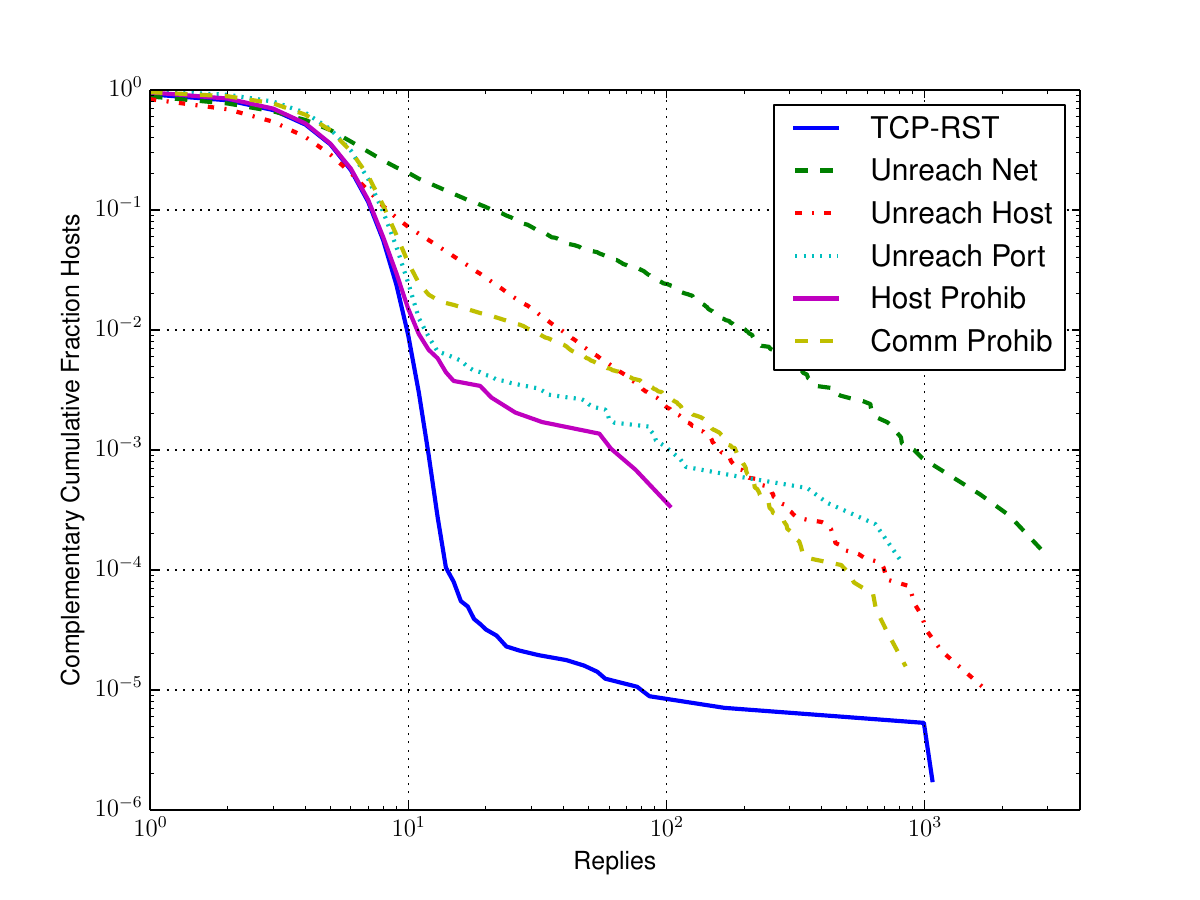}}
  \vspace{-3mm}
  \caption{Number and type of non-TTL exceeded replies received per-host
  during large \yrp probing run.}
  \label{fig:blowback}
  \vspace{-3mm}
\end{figure}

\added{Finally, in consideration of \yrp's stateless TCP probing, we
examine the number and types of replies received during our probing
run.  Figure~\ref{fig:blowback} displays the complementary cumulative
distribution of hosts sending one or more non-TTL exceeded replies.
We received approximately 1.2M TCP RST packets.  99.1\% of the hosts
that send a TCP RST packet sent 10 or fewer, indicating that these
hosts received 10 or fewer probes.  We received $\approx$95K host
unreachable, and $\approx$63k communication prohibited ICMP messages.
A very small number of hosts sent thousands of TCP RST packets;
three three IP addresses within Wanadoo French Telecom send
the majority of all RSTs.  
As \yrp never sends more than 32 probes toward a given destination,
a single IP sending a large number of RSTs is suggestive of a 
middlebox.}

\subsection{Short-Lived Dynamics}
%

As an application of rapid topology discovery, we collect
topology snapshots in rapid succession and analyze their properties
and differences in this subsection.

We gather 67,045 target destinations from CAIDA's May 1, 2016 topology
probing from their San Diego monitor.  Again using the east coast
university vantage point, we run \yrp to probe TTLs 1-32 for these
same 67,045 targets.  We run \yrp at $\approx 4000pps$ and invoke \yrp
three times in succession with a minute pause in-between.  In this
way, each snapshot takes approximately 8 minutes to gather, and each
is separated by a minute.  \added{The permutation key is the same
for each snapshot, thus the random probing order is identical for
each.}  We term the snapshots $S_1, S_2$ and $S_3$ in
chronological order.

The interface-level graph resulting from $S_1$ contains 39,968
interfaces and 46,721 edges, while $S_2$ has 40,038 interfaces and
$46,773$ edges.  $S_3$ contains $39,994$ interfaces and $46,749$
edges.
To better understand the differences between snapshots, we perform a
per-target path comparison.  For each target in \replaced{$S_1$}{$S_i$}, we compare the
discovered path in \replaced{$S_1$}{$S_i$} to the\deleted{discovered} path to that same target in
\replaced{$S_2$}{$S_j$}.  We use the Levenshtein edit distance to measure the per-target
path differences between snapshots.  The edit distance is the minimum number of edits
(insertions, substitutions, or deletions of router interfaces).   

Note that inter-snapshot differences are not attributable to
per-flow load balancing as \yrp keeps the packet header
fields which are used for load balancing constant for the same
destination between snapshots (\S\ref{sec:design:stateless}).

Additionally, to better understand the types of path changes, we
count the frequency of each edit operation and missing hop
substitutions.
These missing hop operations are instances where the
path contains a responsive router for a particular TTL for one
snapshot, but no response at that TTL when probing the same
destination in a subsequent snapshot.  Such missing hops
may be attributable to routers performing ICMP rate
limiting\added{\footnote{While \yrp sends TCP probes, many routers
limit the rate of ICMP responses they return.}}, or may
be due to packet loss. 

%
%

A deeper analysis of the most frequent missing hops between $S_1$ and
$S_2$ reveals that the large majority (92.2\%)
come from the first four hops within the local network of the vantage
point.  Specifically, 73\% of the missing hops are due to the router
at TTL 3, 18\% are due to the router at TTL 1, and 1\% are due to the
router at TTL 4.  In contrast, the router at TTL 2 always responds,
suggest that some of the local routers implement ICMP rate limiting
while one does not. 

\begin{figure}[!t]
  \centering
  \resizebox{0.9\columnwidth}{!}{\includegraphics{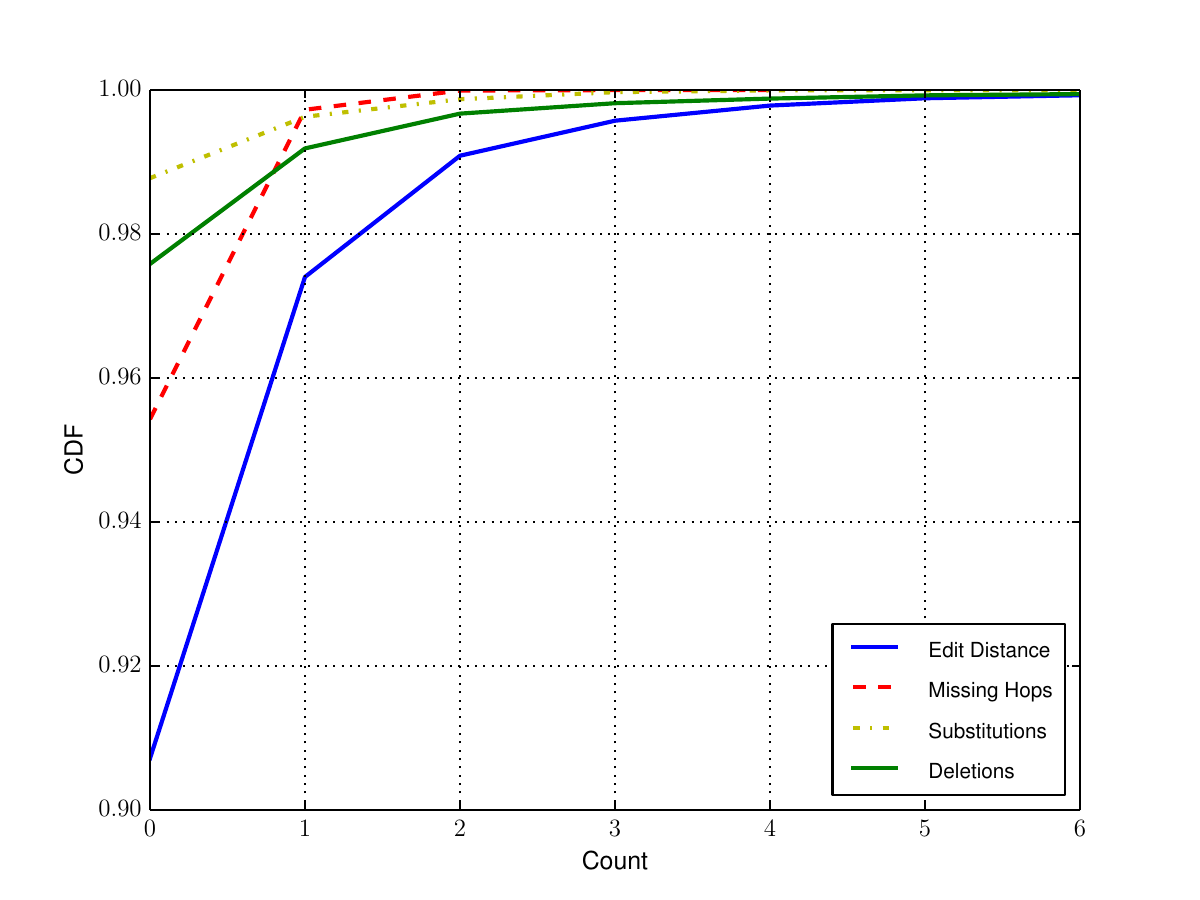}}
  \vspace{-3mm}
  \caption{Cumulative fraction of measured paths versus pair-wise path edit distance 
   between successive \yrp topology snapshots.}
  \label{fig:ed}
  \vspace{-3mm}
\end{figure}

Figure~\ref{fig:ed} displays the results of the edit distance
comparison between $S_1$ and $S_2$, ignoring differences
due to the \added{vantage point's} local network (TTL $\le 4$, as described above).
The paths to approximately 91\% of the destinations are identical
between $S_1$ and $S_2$, while approximately 6\% have a single hop
difference.  Less than 1\% of the destinations show a difference of 
$\ge 2$ hop edits.  
Separated by the edit operation, \deleted{we see that }$\approx$4\% of the 
\added{paths have} 1 hop
differences \added{that} are due to missing hops, 1\% are hop deletions, and
fewer than 1\% are substitutions.

\deleted{Figure~\ref{fig:samplepath}
shows, for each of the three snapshots, the final four responsive hops
toward the destination 188.32.230.138 (in AS 42610).  In the
intermediate snapshot, $S_2$, we see two hop substitutions where the
next hop after AS 1273 (213.185.219.106, Cable and Wireless) changes
to a different sequence of hops within AS 12389 (Rostelecom).  By the
time of the third snapshot, $S_3$, the path changes back to that seen
in $S_1$. 



\begin{figure}[!t]
\tiny{
\begin{Verbatim}[frame=single]
S1: 213.185.219.106 95.167.95.110 94.25.8.173 77.37.254.198
S2: 213.185.219.106 87.226.133.50 188.254.78.129 77.37.254.198
S3: 213.185.219.106 95.167.95.110 94.25.8.173 77.37.254.198
\end{Verbatim}
}
\vspace{-5mm}
\caption{Example of short-lived path dynamics 
observed via successive, rapidly collected topology snapshots.}
\label{fig:samplepath}
\vspace{-2mm}
\end{figure}
}

\added{
To understand the potential of rapidly collected topology
snapshots, we manually investigate and highlight a path exhibiting
a significant change between snapshots. Figure~\ref{fig:samplepath}
shows, for each of the three snapshots, entirely different paths
toward the destination 131.221.200.245 (in AS 262316).  In $S_1$,
the trace leaves our vantage point's local network via AS3356
(Level 3), via AS10578 in $S_2$ (Internet 2 Northeast Gigapop), 
and AS174 (Cogent) in $S_3$.  We manually confirm significant
routing churn for the target's prefix (131.221.200.0/22) evident
in BGP updates archived by Routeviews~\cite{routeviews}.  During
our snapshot collection, there were 176 BGP updates involving
the target prefix, while there were no BGP updates for the same
prefix in the prior 15-minute Routeviews BGP archive.  
}

\added{
Similarly, we find short-lived dynamics within the
core of the network in our snapshots. Figure~\ref{fig:samplepath2}
shows traces toward 129.232.142.175 (AS37153, in South Africa)
traversing Level 3 (AS3356) and AS36351 before reaching AS37153 in
$S_1$, then converging on a path via AS37179 rather than AS36351.
Again, there are no BGP updates for 129.232.128.0/17 in the Routeviews 
archive 15
minutes prior, but 217 updates during, our probing.}

While the exact cause of this short-lived routing change is unknown,
the key point is that it would not have been discovered by the
existing topology mapping systems.  \added{We further find intra-AS dynamics
between paths to the same
destination in our snapshots.  These differences cannot be validated
against available BGP updates visible at Routeviews, but
suggest that \yrp's data-plane inferences may be complementary to 
techniques that rely on the visibility of dynamics within the
control-plane.}
We leave a comprehensive analysis
of the extent and duration of these short-lived dynamics exposed by
\yrp to future work.

\begin{figure}[!t]
\tiny{
\begin{Verbatim}[frame=single]
S1: .. 18.192.9.2 4.53.48.97 4.69.144.80 4.26.0.166 201.48.50.161
S2: .. 18.192.9.2 207.210.142.229 198.71.47.57 * 67.16.148.6 201.48.50.161 
S3: .. 18.192.9.2 38.104.186.185 154.54.30.41 154.54.47.30 154.54.11.110
\end{Verbatim}
}
\vspace{-5mm}
\caption{Example of short-lived path dynamics 
observed near our vantage point via successive, rapidly collected topology snapshots.}
\label{fig:samplepath}
\vspace{-2mm}
\end{figure}

\section{Conclusions}
\label{sec:conclusions}

\yrp demonstrates a new technique for Internet-scale active
\deleted{topology} probing that permits rapid collection of topology
snapshots.  \replaced{O}{As with our initial investigation of short-lived
topology dynamics, o}ur hope is that \yrp facilitates analyses 
not previously possible.  \added{For instance, \yrp can enable a detailed 
longitudinal analysis of short-lived topology dynamics across the
entire Internet in future work.}

\yrp is stable and the code is publicly available~\cite{yarrp}.  
That said, there are several enhancements that would 
be valuable additions.  First, \yrp currently 
only supports IPv4 probing.  Given the vastly larger
IPv6 address space, and relative topological sparsity~\cite{rohrer16},
adding IPv6 support to \yrp could enable more complete
maps of the IPv6 topology to be gathered.  
\added{While the streamlined IPv6 headers prevent direct application of
\yrp's IPv4 header encoding, the full packet quotation in ICMP6
permits more flexibility in recovering state and using different
transport protocols.}

Second, \replaced{while \yrp sends TCP probes, it}{\yrp is currently capable of sending only TCP probes.
It} is well-known that using different transport protocols yields
different responses, due to \deleted{widespread }security and policy
filtering~\cite{keys13}.  \replaced{Adding}{We plan to add} ICMP and UDP probing
to \yrp\deleted{, which} requires utilizing different transport header
fields to encode probe information\added{, while maintaining the first
four bytes constant to keep packets on a single load-balanced path.}
\added{For UDP we expect to encode time into the length and checksum
fields and include a payload that makes the checksum correct.}
\added{For ICMP echo, we can encode the timestamp into the
identifier and sequence number headers, but must include a payload that
produces the same checksum for every packet toward a target.}
\deleted{Doing so is non-trivial
as we must maintain both the Paris-traceroute property of
keeping certain fields constant to keep packets on a single
load-balanced path, while also retaining \yrp's stateless behavior.}

Third, \yrp's stateless and asynchronous nature implies that a
malicious actor could attempt to send bogus responses, while
middleboxes are known to mangle packet
headers~\cite{Craven:2014:MTN:2619239.2626321,malone07}.  In the
future, we wish
to use a keyed cryptographic integrity function over multiple probe
values. Instead of a simple checksum on the target IP address, we
will populate the source port with the value of this keyed integrity
check.
\yrp can then ensure that it both sent the original probe, and that
the probe was not modified in-flight. \deleted{such that the response is not
useful.}

\added{Fourth, while the headers used for load-balancing remain
consistent for all probes toward a given destination, the ICMP
responses that \yrp elicits will have different checksums due to the
quotation containing \yrp's timestamp.  When a router with equal-cost
paths back to the source must generate an ICMP response, it may choose
a source interface based on the ICMP headers (including the ICMP
checksum).  Thus, a load-balancing router at a particular hop may
respond with a \emph{different} IP address between subsequent \yrp
traces.  In future work we plan to address this subtlety.}

Finally, an attractive feature of \yrp's design is the
ability to easily randomize and distribute the probing to multiple
vantage points with negligible coordination and communication overhead.
Similar to the rapid scanning worm
envisioned by Staniford et al.,
the permuted domain can be
distributed~\cite{staniford2002own}.  
\replaced{While the entire domain can be
subdivided among vantage points, doing so causes different vantage
points to probe different TTLs for a given target.  Instead, a 
simple scheme can distribute the probing while ensuring that all
hops toward a target are probed from a consistent source.
Each of $n$ vantage points
permutes the same domain $D = IPs\times TTLs$.  However, the 
$i$'th vantage point only sends a probe for addresses where 
$IP \% (n-1) == i$.  For additional randomness, the IP may be
hashed prior to this check.}
{For $n$ vantage points, and a
given domain $D = IPs\times TTLs$, each vantage point encrypts
$\frac{|D|}{n}$ of the range $i=0,1,\ldots,|D|-1$.  Thus, the $v$'th
vantage point encrypts the range of values $\frac{|D|}{n}v$ to
$\frac{|D|}{n}\left(v+1\right)$ to obtain its sequence of \\
$\langle target IP, TTL\rangle$ values to probe from the overall
permutation.}
The potential speed improvement 
is linearly proportional then to the number of vantage points. 
Only the values $|D|$, $i$\added{, and $key$} need be sent to each vantage
point to distribute the permuted space and achieve complete randomized 
coverage.  Given our empirical (and conservative)
$100kpps$ \yrp rate in this work, we estimate that it is possible to
implement a distributed \yrp among $\approx$128 vantage points to 
traceroute to every routed IPv4 address ($\approx 2^{31}$ targets) in approximately one hour.
\yrp may thus facilitate rapid collection of \emph{complete} Internet
snapshots in the future.

\begin{figure}[!t]
\tiny{
\begin{Verbatim}[frame=single]
S1: .. 4.69.166.5 212.113.14.82 50.97.19.43 5.10.118.137 159.8.138.4
S2: .. 4.69.166.5 4.69.167.82 50.97.19.43 41.84.12.81 41.66.132.246
S3: .. 4.69.166.5 4.69.167.82 212.187.195.2 41.84.12.81 41.66.132.246
\end{Verbatim}
}
\vspace{-5mm}
\caption{Example of path dynamics observed within the core of the
network.}
\label{fig:samplepath2}
\vspace{-2mm}
\end{figure}

\section*{Acknowledgments}
\added{
We thank Simson Garfinkel and Nick Weaver for initial discussions,
Lance Alt for libcperm, 
Garrett Wollman for network administration, 
and Priya Mahadevan for shepherding.
kc claffy, Ann Cox, Mark Gondree, 
Matthew Luckie, and Justin Rohrer
provided invaluable feedback.  This work supported in part by NSF
grant CNS-1213155.  Views and conclusions are those of the authors and
should not be interpreted as representing the official policies or
position of the U.S.\ government or the NSF.
}

\clearpage
\newpage
\small{
\bibliographystyle{abbrv}
\bibliography{yarrp}

\begin{thebibliography}{10}

\bibitem{routeviews}
University of {O}regon {R}oute{V}iews, 2016.
\newblock \url{http://www.routeviews.org}.

\bibitem{augustin2006avoiding}
B.~Augustin, X.~Cuvellier, B.~Orgogozo, F.~Viger, T.~Friedman, M.~Latapy,
  C.~Magnien, and R.~Teixeira.
\newblock Avoiding traceroute anomalies with {Paris} traceroute.
\newblock In {\em Proceedings of ACM IMC}, pages 153--158, 2006.

\bibitem{rfc1812}
F.~Baker.
\newblock {Requirements for IP Version 4 Routers}.
\newblock RFC 1812 (Proposed Standard), June 1995.

\bibitem{baltra14ips}
G.~Baltra, R.~Beverly, and G.~G. Xie.
\newblock {Ingress Point Spreading: A New Primitive for Adaptive Active Network
  Mapping}.
\newblock In {\em Proceedings of Passive and Active Network Measurement (PAM)},
  pages 56--66, Mar. 2014.

\bibitem{yarrp}
R.~Beverly.
\newblock Yarrp, 2016.
\newblock \url{https://www.cmand.org/yarrp}.

\bibitem{Beverly:2010:PAI:1879141.1879162}
R.~Beverly, A.~Berger, and G.~G. Xie.
\newblock {Primitives for active Internet topology mapping: toward
  high-frequency characterization}.
\newblock In {\em Proceedings of ACM IMC}, pages 165--171, 2010.

\bibitem{ciphers}
J.~Black and P.~Rogaway.
\newblock Ciphers with arbitrary finite domains.
\newblock In {\em Topics in Cryptology--CT-RSA}, pages 114--130. Springer,
  2002.

\bibitem{caida-topo}
{CAIDA}.
\newblock {The CAIDA UCSD IPv4 Routed /24 Topology Dataset}, 2016.
\newblock
  \url{http://www.caida.org/data/active/ipv4_routed_24_topology_dataset.xml}.

\bibitem{Craven:2014:MTN:2619239.2626321}
R.~Craven, R.~Beverly, and M.~Allman.
\newblock {A Middlebox-cooperative TCP for a Non End-to-end Internet}.
\newblock In {\em Proceedings of ACM SIGCOMM}, pages 151--162, 2014.

\bibitem{cunha2016sibyl}
{\'I}.~Cunha, P.~Marchetta, M.~Calder, Y.-C. Chiu, B.~V. Machado,
  A.~Pescap{\`e}, V.~Giotsas, H.~V. Madhyastha, and E.~Katz-Bassett.
\newblock Sibyl: a practical internet route oracle.
\newblock In {\em 13th USENIX Symposium on Networked Systems Design and
  Implementation (NSDI 16)}, pages 325--344, 2016.

\bibitem{cunha2011predicting}
I.~Cunha, R.~Teixeira, D.~Veitch, and C.~Diot.
\newblock Predicting and tracking internet path changes.
\newblock {\em ACM SIGCOMM Computer Communication Review}, 41(4):122--133,
  2011.

\bibitem{donnet2005efficient}
B.~Donnet, P.~Raoult, T.~Friedman, and M.~Crovella.
\newblock Efficient algorithms for large-scale topology discovery.
\newblock {\em ACM SIGMETRICS Performance Evaluation Review}, 33(1):327--338,
  2005.

\bibitem{durumeric2013zmap}
Z.~Durumeric, E.~Wustrow, and J.~A. Halderman.
\newblock Zmap: Fast internet-wide scanning and its security applications.
\newblock In {\em USENIX Security}, pages 605--620, 2013.

\bibitem{graham14}
R.~Graham, P.~McMillan, and D.~Tentler.
\newblock {Mass Scanning the Internet}.
\newblock In {\em {DEF CON 22}}, 2014.

\bibitem{gunes2008resolving}
M.~H. Gunes and K.~Sarac.
\newblock Resolving anonymous routers in {Internet} topology measurement
  studies.
\newblock {\em INFOCOM}, pages 1076--1084, 2008.

\bibitem{huffaker2012topology}
B.~Huffaker, M.~Fomenkov, and k.~claffy.
\newblock Internet topology data comparison.
\newblock {\em Cooperative Association for Internet Data Analysis (CAIDA)},
  2012.

\bibitem{caida-ark}
Y.~Hyun and k.~claffy.
\newblock Archipelago measurement infrastructure, 2015.
\newblock \url{http://www.caida.org/projects/ark/}.

\bibitem{traceroute}
V.~Jacobson.
\newblock traceroute, 1989.
\newblock \url{ftp://ftp.ee.lbl.gov/traceroute.tar.gz}.

\bibitem{claffy09}
k.~claffy, Y.~Hyun, K.~Keys, and M.~Fomenkov.
\newblock Internet mapping: from art to science.
\newblock In {\em Proceedings of IEEE Cybersecurity Applications and
  Technologies Conference for Homeland Security}, Mar. 2009.

\bibitem{keys13}
K.~Keys, Y.~Hyun, M.~Luckie, and k.~claffy.
\newblock {Internet-Scale IPv4 Alias Resolution with MIDAR}.
\newblock {\em Transactions on Networking}, 21(2):383--399, Apr 2013.

\bibitem{luckie10scamper}
M.~Luckie.
\newblock Scamper: a scalable and extensible packet prober for active
  measurement of the {Internet}.
\newblock In {\em IMC}, pages 239--245, Nov. 2010.

\bibitem{iplane06}
H.~V. Madhyastha, T.~Isdal, M.~Piatek, C.~Dixon, T.~Anderson, A.~Krishnamurthy,
  and A.~Venkataramani.
\newblock {iPlane}: An information plane for distributed services.
\newblock In {\em Proceedings of USENIX OSDI}, Nov. 2006.

\bibitem{malone07}
D.~Malone and M.~Luckie.
\newblock Analysis of {ICMP} quotations.
\newblock In {\em Proceedings of the 8th Passive and Active Measurement ({PAM})
  Workshop}, Apr. 2007.

\bibitem{rohrer16}
J.~Rohrer, B.~LaFever, and R.~Beverly.
\newblock {Empirical Study of Router IPv6 Interface Address Distributions}.
\newblock {\em {IEEE Internet Computing}}, July 2016.

\bibitem{sklower1991tree}
K.~Sklower.
\newblock A tree-based packet routing table for berkeley unix.
\newblock In {\em USENIX Winter}, volume 1991, pages 93--99, 1991.

\bibitem{spring2002measuring}
N.~Spring, R.~Mahajan, and D.~Wetherall.
\newblock Measuring {ISP} topologies with {Rocketfuel}.
\newblock {\em ACM SIGCOMM Computer Communication Review}, 32(4):133--145,
  2002.

\bibitem{staniford2002own}
S.~Staniford, V.~Paxson, N.~Weaver, et~al.
\newblock How to own the internet in your spare time.
\newblock In {\em USENIX Security}, pages 149--167, 2002.

\bibitem{rfc6864}
J.~Touch.
\newblock {Updated Specification of the IPv4 ID Field}.
\newblock RFC 6864 (Proposed Standard), Feb. 2013.

\bibitem{mathinternet}
W.~Willinger, D.~Alderson, and J.~C. Doyle.
\newblock Mathematics and the {Internet}: A source of enormous confusion and
  great potential.
\newblock {\em Notices of the AMS}, 56(5), 2009.

\end{thebibliography}
}

\end{document}